\documentclass[%
 reprint,
superscriptaddress,
 amsmath,amssymb,
 aps,
 prl,
floatfix
]{revtex4-1}
\usepackage{pdfpages}
\usepackage{pgffor}
\usepackage{graphicx}
\usepackage{dcolumn}
\usepackage{bm}
\usepackage{hyperref}
\usepackage{xr}
\usepackage{physics}
\usepackage{chemformula}
\usepackage{siunitx}
\usepackage{color,soul}

\newcommand{\anni}{\hat{c}}
\newcommand{\crea}{\hat{c}^{\dagger}}
\newcommand{\cpkp}{c{'}\mathbf{k{'}}}
\newcommand{\vpkp}{v{'}\mathbf{k{'}}}
\newcommand{\bmk}{\mathbf{k}}
\newcommand{\bmkp}{\mathbf{k{'}}}
\newcommand{\bmq}{\mathbf{q}}
\renewcommand{\braket}[1]{\expval{#1}}

\makeatletter
\AtBeginDocument{\let\LS@rot\@undefined}
\makeatother

\begin{document}

\preprint{APS/123-QED}

    \title{Phonon-Assisted Ballistic Current From First Principles Calculations}

    \author{Zhenbang Dai}
    \thanks{These authors contributed equally}
	\affiliation{Department of Chemistry, University of Pennsylvania, Philadelphia, Pennsylvania 19104--6323, USA}
    \author{Aaron M. Schankler}%
    \thanks{These authors contributed equally}
	\affiliation{Department of Chemistry, University of Pennsylvania, Philadelphia, Pennsylvania 19104--6323, USA}
    \author{Lingyuan Gao}%
	\affiliation{Department of Chemistry, University of Pennsylvania, Philadelphia, Pennsylvania 19104--6323, USA}
    \author{Liang Z. Tan}%
	\affiliation{Molecular Foundry, Lawrence Berkeley National Laboratory, Berkeley, California 94720, USA}
    \author{Andrew M. Rappe}%
	\affiliation{Department of Chemistry, University of Pennsylvania, Philadelphia, Pennsylvania 19104--6323, USA}

\date{\today}

\begin{abstract}
The bulk photovoltaic effect (BPVE) refers to  current generation due to illumination by light in a homogeneous bulk material lacking inversion symmetry. In addition to the intensively studied shift current, the ballistic current, which originates from asymmetric carrier generation due to scattering processes, also constitutes an important contribution to the overall kinetic model of the BPVE. In this letter, we use a perturbative approach to derive a formula for the ballisic current resulting from the intrinsic electron-phonon scattering in a form amenable to  first-principles calculation. We then implement the theory and calculate the ballistic current of the prototypical BPVE material \ch{BaTiO3} using quantum-mechanical density functional theory. The magnitude of the ballistic current is comparable to that of shift current, and the total spectrum (shift plus ballistic) agrees well with the experimentally measured photocurrents.
Furthermore, we show that the ballistic current is sensitive to structural change, which could benefit future photovoltaic materials design.

\begin{description}
\item[Keywords]
BPVE, shift current, ballistic current, first principle, electron-phonon coupling
\end{description}
\end{abstract}

\maketitle

The bulk photovoltaic effect (BPVE) is the phenomenon of photocurrent generation in a homogeneous material that lacks inversion symmetry \cite{belinicher1980,VonBaltz1981}.  Compared to traditional photovoltaic devices with a p-n junction to separate electron-hole pairs, where the power conversion efficiency cannot go beyond the Shockley-Queisser limit \cite{shockley1961}, the BPVE can generate large short-circuit photocurrent and above-bandgap photovoltage, thus potentially surpassing the efficiency limit of conventional solar cells \cite{tan2016shift,spanier2016}.



Shift current, which is a purely quantum mechanical effect, is considered to be one of the dominant mechanisms of the BPVE. The shift current results from the coherent evolution of a quantum wave packet; a net current is generated by a real-space shift of excited electrons under illumination. The shift current has been extensively studied analytically and is also readily obtained from first-principles calculations based on electronic structure \cite{sipe2000sc,young2012first,ibanez2018wfshift,ibanez2018wfshift,wang2017first,morimoto2016topological}. This enables \emph{ab-initio} study of the shift current response of a wide variety of materials, including\cite{zhang2019strong,ibanez2020directional,brehm2018predicted,rangel2017large}. Though no overarching design rules have been established, previous studies have established links between shift current response and wavefunction delocalization and polarization \cite{tan2016shift,tan2019limit,fregoso2017polarization,cook2017design}. Although the shift current mechanism is a major component of the BPVE, our recent first principles study shows that it cannot fully account for the experimental photocurrent spectrum of \ch{BaTiO3} \cite{fei2020shift}. Indeed, unlike shift current which is a purely excitation theory, kinetic processes including the relaxation and recombination of photo-excited carriers are often not taken into account. Therefore, other mechanisms related with kinetic processes must also be studied for a full understanding of the BPVE.





Ballistic current, which is a current based on carrier transport, results from asymmetric occupation of carriers at momentum  $\mathbf{k}$ and $-\mathbf{k}$ \cite{belinicher1988relation}, and it is viewed as a dominant mechanism for the BPVE by \cite{Sturman_2019,alperovich1982photogalvanic,duc2019ballistic,Shelest1979ballistic}. In the absence of inversion symmetry, the occupation is determined by different asymmetric scattering processes, including scattering from defects, electron-hole interactions, and the electron-phonon interactions \cite{belinicher1988relation, belinicher1978phonon,alperovich1982photogalvanic,gu2017mesoscopic}, whereas for magnetic systems which break time-reversal symmetry, the asymmetric momentum distribution can still exist without these scattering mechanisms\cite{belinicher1980,alperovich1982photogalvanic,zhang2019switchable}. We will focus on materials with time-reversal symmetry. Among these asymmetric scattering processes, electron-phonon scattering is an intrinsic mechanism present regardless of the quality of the material, and it will be strongly influenced by temperature. As revealed in \cite{burger2019direct}, both ballistic current and shift current are significant mechanisms for BPVE in \ch{Bi12GeO20}. Although there are several previous studies calculating ballistic current, they are based on few-band models, and approximations are usually made assuming that only certain regions of the Brillouin zone contribute \cite{alperovich1982photogalvanic,belinicher1978phonon,Shelest1979ballistic,duc2019ballistic}. Therefore, to establish the importance of ballistic current for real materials, it is imperative to have a study based on the full electronic structure and phonon dispersion.

In this Letter, we perform a first-principles study of the ballistic current due to electron-phonon scattering (referred to here simply as the ballistic current). To the best of our knowledge, no such calculation has yet been reported. Following previous work \cite{belinicher1988relation,belinicher1978phonon}, we take the electron-phonon coupling as the source of scattering and derive the asymmetric carrier generation rate using a Kubo formula. With the developed {\em ab-initio} Fr\"olich electron-phonon interaction, the carrier generation rate can be calculated
in an \emph{ab-initio} way. With the rate and band velocities, current can be calculated according to the Boltzmann transport equation. We compare our results with theoretically calculated shift current and also with the experimentally measured photocurrent of \ch{BaTiO3} \cite{koch1975bulk,fei2020shift}. We also explore the ballistic current in different crystal structures of this material, and we find that the magnitude of ballistic current can vary significantly.

\begin{figure}
\includegraphics[width=85mm]{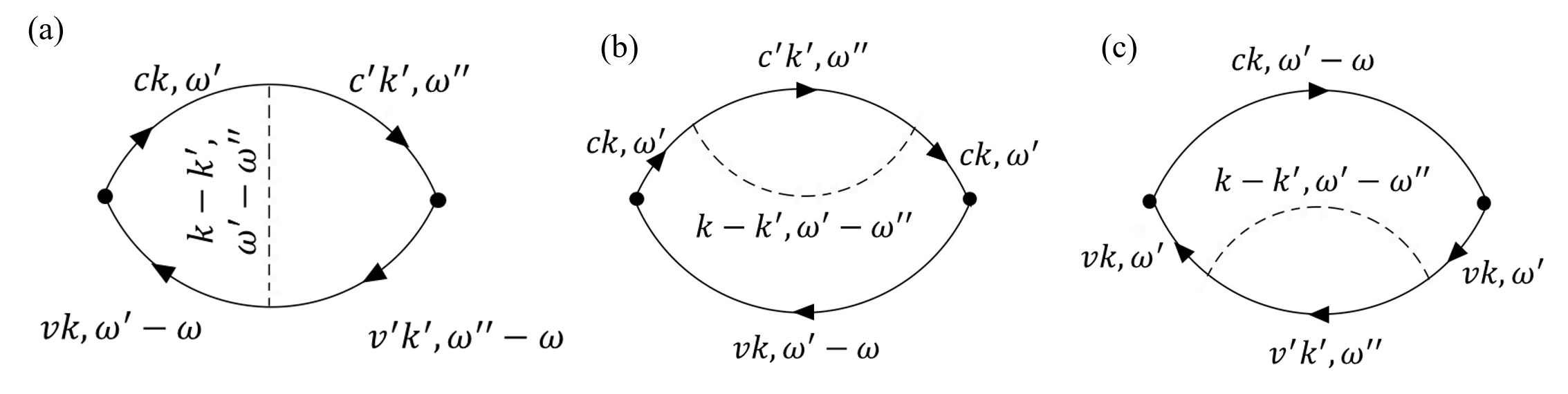}
\caption{\label{fig:feynman} Three different Feynman diagrams for the second-order expansion of the momentum-momentum correlation function with the electron-phonon coupling as the perturbation. Only diagram (a) will contribute to asymmetric scattering. }
\end{figure}


Based on the Boltzmann transport equation, the phonon-mechanism ballistic current can be expressed as:
\begin{align}
\label{eqn:8}
    j^{\alpha\beta,\gamma}(\omega)
    &= 2e\tau_0 \sum_{cv\bmk} \Gamma_{cv,\bmk}^{\alpha\beta,asym}(\omega) \bqty{ v_{c\bmk}^{e,\gamma} -
    v_{v\bmk}^{e,\gamma}}, \nonumber \\
\end{align}
where $\Gamma_{cv,\bmk}^{\alpha\beta,asym}(\omega)=\frac{1}{2}\big(\Gamma_{cv,\bmk}^{\alpha\beta}(\omega)-\Gamma_{cv,-\bmk}^{\alpha\beta}(\omega)\big)$ is the asymmetric carrier generation rate for an electron-hole pair $(c,v)$ at $\mathbf{k}$, $e$ is the electron charge, $\tau_0$ is the momentum relaxation time, and $\mathbf{v}^e_{c\bmk}$ ($\mathbf{v}^e_{v\bmk}$) is the electron (hole) velocity obtained from band derivatives. The leading factor of two is for spin degeneracy. The central quantity that needs to be evaluated is the asymmetric carrier generation rate, and it is derived below.

Adopting the velocity gauge $\mathbf{E} = -\frac{\partial \mathbf{A}}{\partial t}$ and taking the electron-photon interaction as $\hat{H}_{e-photon} = \frac{e}{m}\mathbf{\hat{P}}\cdot\mathbf{\hat{A}}$, from linear response theory \cite{jishi2013feynman,giuliani2005quantum}, the average power delivered by monochromatic light of frequency $\omega$ to the system during one period of oscillation is
\begin{equation}
    W = -2\omega \Im\bqty{\chi^{\alpha\beta}(\omega)}
    \Big( {\frac{e}{m\omega}} \Big)^2 
    E_{\alpha}(\omega)E_{\beta}(\omega) ,
\end{equation}
where ${\chi}^{\alpha\beta}(\omega)$ is the rank-two response function in the presence of $\mathbf{E}$ field with  Greek letters denoting its components, $e$ and $m$ are the electron charge and mass, and $E_{\alpha}(\omega)$ is the amplitude of the electric field, whose frequency dependence will be taken implicitly hereafter. Considering that each photon absorbed will be converted to an electron and hole, \cite{gu2017mesoscopic}, the overall carrier generation rate $\Gamma^{\alpha\beta}(\omega)$ can then be written as 
\begin{equation}
\label{eqn:2}
    \Gamma^{\alpha\beta}(\omega) = \frac{W}{{\hbar}\omega} =
    -\frac{2}{\hbar}
    \Im\bqty{\chi^{\alpha\beta}(\omega)}
    \Big( {\frac{e}{m\omega}} \Big)^2 
    E_{\alpha}E_{\beta}.
\end{equation}
According to the Kubo formula, the response function is related to the retarded momentum-momentum correlation function:
\begin{align}
\label{eqn:3}
    {\chi}^{\alpha\beta}(\omega)
    &=\frac{1}{\hbar}C^{R}_{\hat{P}^{\alpha}{\hat{P}}^{\dagger\beta}}(\omega) \nonumber \\
    &= -\frac{i}{{\hbar}} \int_{-\infty}^{+\infty} dt e^{-i{\omega}t} {\Theta(t)} \braket{\comm{\hat{P}^{\alpha}(t)}{{\hat{P}}^{\dagger\beta}(0)}}.
\end{align}
 Here, the brackets $\braket{ \cdot  }$ indicate an equilibrium average with respect to the total Hamiltonian that includes any extra interaction $\hat{H}'$, which in our case is the electron-phonon interaction, and the momentum operators are in the Heisenberg picture. To evaluate $\chi^{\alpha\beta}(\omega)$, we first calculate the imaginary-time (Matsubara) correlation function in its second quantization form with Bloch states as the basis:
\begin{multline}
\label{eqn:4}
    {\chi_T^{\alpha\beta}}(i\omega_n) = -\frac{1}{\hbar} \sum_{\bmk\bmkp cc'vv'}
    \braket{v\bmk|\hat{P}^{\alpha}|c\bmk}\braket{c{'}\bmkp|\hat{P}^{\beta}|v{'}\bmkp} \\
    \times
    \int_{0}^{\hbar/k_{B}T} d{\tau} e^{i{\omega_n}{\tau}}
    \braket{\hat{T}_{\tau} {\crea_{v\bmk}(\tau)} {\anni_{c\bmk}(\tau)} {\crea_{\cpkp}(0)} {\anni_{\vpkp}(0)} }, 
\end{multline}
where $c (c')$ and $v (v')$ are band indices for conduction and valence bands, respectively, $\mathbf{k}, \mathbf{k}'$ are crystal momenta, and $1/k_BT$ reflects the influence of temperature\cite{mahan2013many}.
The retarded and Matsubara correlation functions can be related through analytical continuation: ${{\chi}^{\alpha\beta}}(\omega)={{\chi}_{T}^{\alpha\beta}}({i\omega_n\xrightarrow[]{}\omega+i0^{+}})$, where $0^{+}$ is a infinitesimal positive number. In Eq.~\ref{eqn:4}, two conditions hold: first, due to Pauli exclusion, transitions are only allowed from occupied valence bands to unoccupied conduction bands; also the population of electrons in a semiconductor is not significantly influenced by temperature.
Second, because of the negligible momentum carried by photons, only vertical transitions are allowed. From Eq.~\ref{eqn:4}, it can be seen that the carrier generation rate $\Gamma^{\alpha\beta}(\omega)$ can be decomposed into components from various $\mathbf{k}$ points and electron-hole pairs: $\Gamma^{\alpha\beta}(\omega) = \sum_{cv\bmk}\Gamma_{cv,\bmk}^{\alpha\beta}(\omega)$, and we only consider the asymmetric scatterings $\Gamma_{cv,\bmk}^{\alpha\beta}(\omega) \neq \Gamma_{cv,-\bmk}^{\alpha\beta}(\omega)$ as the contribution to net current. Without any other interaction, Eq.~\ref{eqn:4} corresponds to Fermi's golden rule, and this is a symmetric excitation which does not generate any current. 

Therefore, we calculate the carrier generation rate in the presence of electron-phonon coupling, which will impose the influence of temperature. By introducing the Fr\"olich e-ph Hamiltonian as\cite{frohlich1950theory,jishi2013feynman,mahan2013many}
\begin{equation}
\label{eqn:5}
    \hat{H}'_{e-phonon}
    = \sum_{{\mu}nn{'}} \sum_{\bmk\bmkp} g_{{\mu}\bmk\bmkp}^{nn{'}} 
    {\crea_{n{'}\bmkp}} {\anni_{n\bmk}} {\hat{\Phi}_{\bmk-\bmkp}^{\mu}}
\end{equation}
where $\hat{\Phi}_{\mathbf{q}}^{\mu}=\hat{a}_{\mathbf{q}}^{\mu}+\hat{a}_{-\mathbf{q}}^{\mu\dagger}$ is the phonon field operator, $\hat{a}_{\mathbf{q}}^{\mu}$($\hat{a}_{\mathbf{q}}^{\mu\dagger}$) are the phonon annihilation(creation) operators, and $g_{{\mu}\bmk\bmkp}^{nn{'}}$ is the electron-phonon coupling matrix, we perform a perturbative expansion using a Feynman diagrammatic approach. The lowest-order non-zero contribution is second-order, illustrated as three different diagrams in Fig.~\ref{fig:feynman}). As shown in the Supplementary Material, the processes of Fig.~\ref{fig:feynman}b and \ref{fig:feynman}c are symmetric scattering, and only Fig.~\ref{fig:feynman}a contributes to asymmetric scattering. By applying Feynman rules on Fig.~\ref{fig:feynman}a and performing analytical continuation, we can find the second-order correction to the carrier generation rate $\Delta\Gamma_{cv,\bmk}^{\alpha\beta}(\omega)$. Finally, we use relations that are satisfied for materials with time-reversal symmetry
\begin{align}
\braket{v,-\bmk|\hat{P}^{\alpha}|c,-\bmk}&=-\braket{v,\bmk|\hat{P}^{\alpha}|c,\bmk}^{*} \nonumber \\
g_{{\mu}-\bmk-\bmkp}^{nn{'}} &= \pqty{g_{{\mu}\bmk\bmkp}^{nn'}}^*
\end{align}
to write the asymmetric carrier generation rate:



\begin{widetext}
\begin{align}
\label{eqn:7}
    &\Gamma_{cv,\bmk}^{\alpha\beta, asym}(\omega)\nonumber \\
    &=\frac{1}{2}\Big( \Delta\Gamma_{cv,\bmk}^{\alpha\beta}(\omega)-\Delta\Gamma_{cv,-\bmk}^{\alpha\beta}(\omega) \Big)
    =\frac{2}{\hbar}
    {\bigg( {\frac{{\pi}e}{{m\omega}}} \bigg)}^2
    E_{\alpha}E_{\beta}
    \sum_{c{'}v{'}\bmkp\mu}
    Im\big[\braket{v\bmk|\hat{P}^{\alpha}|c\bmk}\braket{c{'}\bmkp|\hat{P}^{\beta}|v{'}\bmkp}
    g_{{\mu}\bmk\bmkp}^{cc{'}}g_{{\mu}\bmkp\bmk}^{v{'}v}
    \big]\nonumber\\
    &\times
    \bigg\{(N_{\bmq}+1)
    \bigg[ \delta(E_{c\bmk}-E_{v\bmk}-\hbar \omega)
    \delta(E_{\cpkp}-E_{\vpkp}-\hbar \omega)
    \Big( \pv{\frac{1}{E_{\cpkp} - E_{c\bmk} + \hbar \omega_{\bmq}}} + 
    \pv{\frac{1}{E_{v\bmk} - E_{\vpkp} + \hbar \omega_{\bmq}}} 
    \Big) \nonumber\\
    &+\delta(E_{c\bmk}-E_{v\bmk}-\hbar \omega)
    \pv{\frac{1}{E_{\cpkp}-E_{\vpkp}-\hbar \omega}}
    \Big(
    \delta{(E_{\cpkp} - E_{c\bmk} + \hbar \omega_{\bmq})} + 
    \delta{(E_{v\bmk} - E_{\vpkp} + \hbar \omega_{\bmq}}) 
    \Big)\nonumber\\
    &+\pv{\frac{1}{E_{c\bmk}-E_{v\bmk}-\hbar \omega}}
    \delta(E_{\cpkp}-E_{\vpkp}-\hbar \omega)
    \Big(
    \delta{(E_{c\bmk} - E_{\cpkp} + \hbar \omega_{\bmq})} + 
    \delta{(E_{\vpkp} - E_{v\bmk} + \hbar \omega_{\bmq}}) \Big) 
    \bigg]+N_{\bmq}[\omega_{\bmq}\Leftrightarrow-\omega_{\bmq}]
    \bigg\}
\end{align}
\end{widetext}
where $\mathbf{q = k-k'}$ is the phonon momentum, $N_{\bmq}$ is the phonon population, and $[\omega_\bmq \Leftrightarrow -\omega_{\bmq}]$ denotes the term in brackets in Eq.~\ref{eqn:7} with instances of $\omega_{\bmq}$ negated, $\omega_{\bmq}$ being the phonon dispersion. 
The delta functions in Eq.~\ref{eqn:7} reflect the selection rule for optical transition, and the electron-phonon coupling matrices together with principal parts are the modulation of the transition rate. 
The initial asymmetric carrier distribution quickly thermalizes, so the carriers contribute to the current only for times on the order of the momentum relaxation time of the carriers, which is usually on the femtosecond time scale~\cite{gu2017mesoscopic,burger2019direct}. We approximate $\tau_0$ to be \SI{2}{\femto\second} in this work, which is justified by an estimation from first-principles calculations (See SI). Together, Eq.~\ref{eqn:8} and Eq.~\ref{eqn:7} provide a method to compute the ballistic current density from quantities that are readily available from first-principles calculations.


We perform density functional theory (DFT) and density functional perturbation theory (DFPT) calculations using the \textsc{Quantum Espresso} package \cite{giannozzi2009quantum, giannozzi2017espresso}. Generalized gradient approximation exchange correlation functional and norm-conserving pseudopotentials produced by the OPIUM package are used \cite{perdew1996pbe, rappe1990optimized, ramer1999designed}. The convergence threshold for self-consistent calculations was \SI{e-8}{Ry/cell}, and for DFPT calculations it was \SI{e-16}{Ry/cell}. Velocity and electron-phonon coupling matrices are calculated by Wannier interpolation using the EPW package \cite{giustino2007epw, ponce2016epw}. All quantities are sampled on an \num{8x8x8} unshifted Monkhorst-Pack grid \cite{monkhorst1976special}, and the principal part integration is dealt with using a generalized Newton-Cotes method (See~SI).

\begin{figure}
\includegraphics[width=80mm]{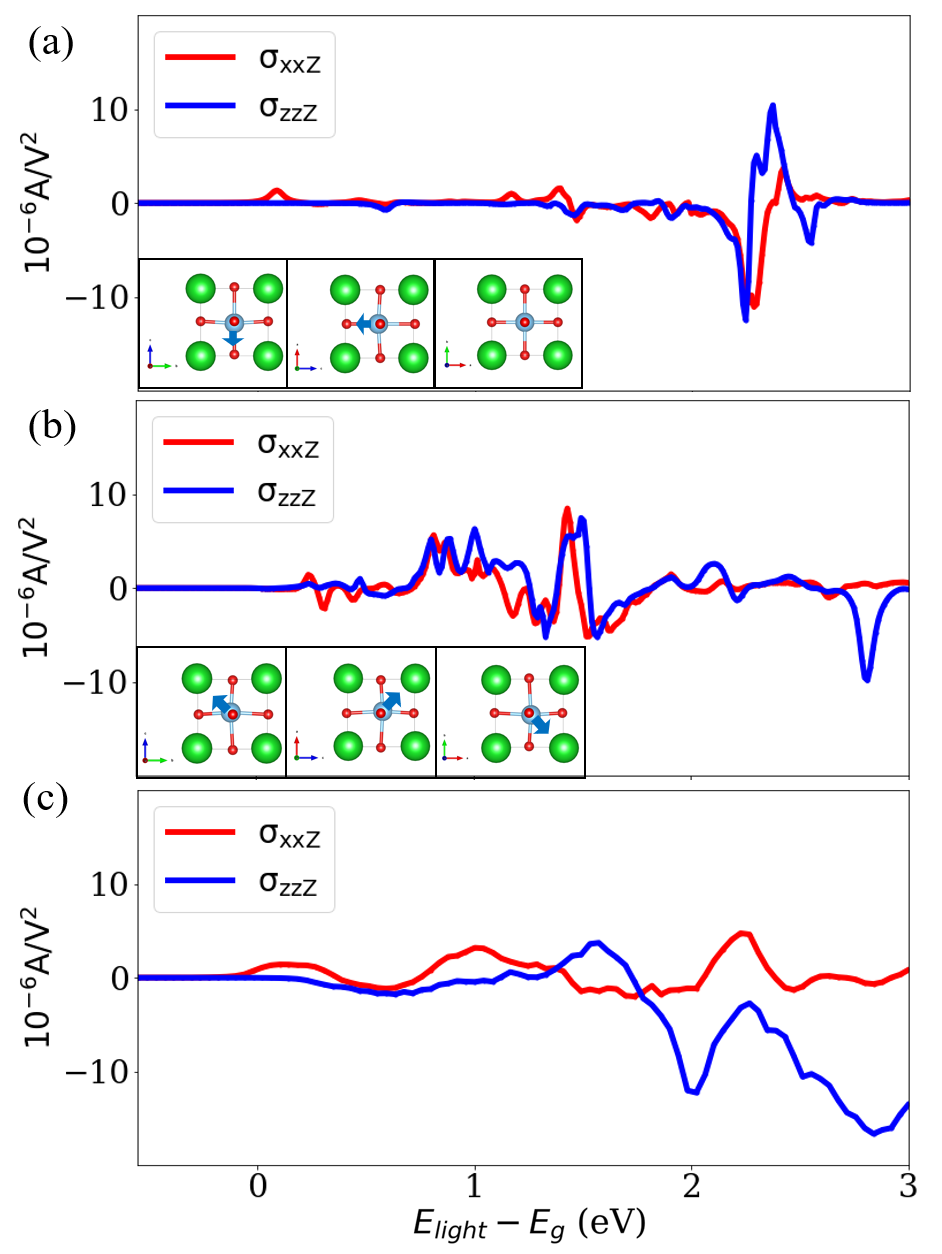}
\caption{\label{fig:response} First-principles results for \ch{BaTiO3}. (a) The ballistic current for the room-temperature tetragonal phase. (b) The ballistic current for the 0~K rhombohedral phase. (c) The shift current for the room-temperature tetragonal phase (reproduced from \cite{young2012first}). The insets of (a) and (b) show the structures of \ch{BaTiO3} for each phase. It can be seen that the ballistic current and the shift current are of similar magnitude, and that structural phase transition in \ch{BaTiO3} can dramatically change the ballistic current response.}
\end{figure}

BaTiO$_3$, as a prototypical ferroelectric and bulk photovoltaic material, is an ideal candidate for benchmarking the ballistic current; the BPVE current spectrum has been measured for BaTiO$_3$~\cite{koch1975bulk}, and the shift current has also been predicted by first-principles calculations.~\cite{young2012first} We use the experimental lattice parameters of tetragonal BaTiO$_3$ with Ti-displacement along (001) to represent the spatially-averaged structure, and the atomic positions are relaxed before the phonon calculations. The temperature of phonon is chosen to be the room temperature. The theoretical ballistic current is shown in Fig.~\ref{fig:response}(a). We find that the ballistic current has a more jagged response profile, which is indicated by \cite{Sturman_2019} as a signature of the ballistic current. For the considered range of light frequency, the largest calculated response occurs at 2.1--2.5~eV above the band gap, similar to the shift current (Fig.~\ref{fig:response}(c)). Even though the lineshape of the ballistic current is more complicated, we note that the turn-on frequency of  $\sigma_{zzZ}$ is larger than that of $\sigma_{xxZ}$ for both ballistic and shift current. In addition, the amplitudes of the ballistic and shift current and similar in magnitude, and thus we find that both shift current and ballistic current will contribute significantly to the experimentally measured current. 

\begin{figure}
\includegraphics[width=85mm]{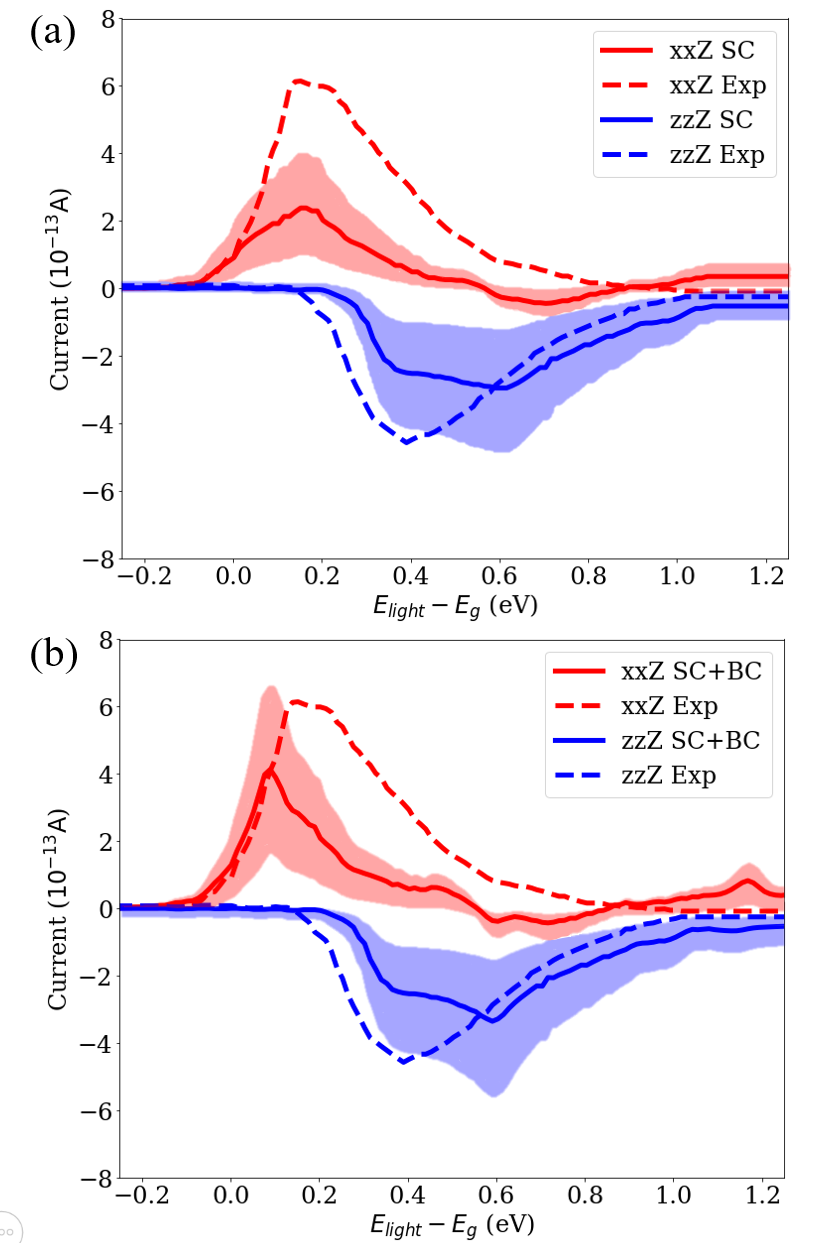}
\caption{\label{fig:exp} Comparison between the theoretical and experimental results for tetragonal \ch{BaTiO3}~\cite{koch1975bulk, fei2020shift}. (a) The comparison between the experimental BPVE and the theoretical shift current (SC, reproduced from \cite{fei2020shift}). (b) The comparison between the experimental BPVE and the theoretical shift current plus  ballistic current (SC+BC). The solid lines are computed by assuming 0.5~$\mathrm{mW/cm^2}$ light intensity and 0.15~$\mathrm{cm}$ sample width.The shaded areas account for the range of experimental parameters in \protect{\cite{koch1975bulk,koch1976anomalous}} that gives the boundary of the response. For the $xxZ$ component, the ballistic current  partially fills the gap between the shift current and experimental spectra. For the $zzZ$ component, the shift current alone agrees fairly closely with experiment, and the ballistic current barely influences the theoretical lineshape.}
\end{figure}

To compare with experiment, we calculate the real photocurrent based on the Glass coefficient\cite{glass_linde_negran_1974,fei2020shift}, by further computing the absorption coefficient with the quasi-particle correction. As pointed out by our previous work~\cite{fei2020shift}, the quasi-particle correction will significantly influence the  absorption profile, but it will mainly blue-shift the response tensor within the frequency range of interest. We apply the same technique by calculating the absorption coefficient using quasi-particle energies with exciton correction while  calculating the current response tensor at the GGA level followed by a rigid shift to account for the underestimation of the band gap (2.1 eV at DFT-GGA level to 3.78 eV at BSE level.\cite{fei2020shift}) In this way, accuracy is improved while the computational cost is kept low.
In addition, we also consider the experimental errors for sample dimensions and light intensities as reported in \cite{young2012first,koch1975bulk,koch1976anomalous}. 
In Fig.~\ref{fig:exp} the $xxZ$ ballistic current partially fills the gap between the shift current and the experimental spectra, whereas for the $zzZ$ component whose shift current has already aligned well with the experiments, the ballistic current barely influence the theoretical BPVE spectrum. This confirms that the ballistic current from the electron-phonon scattering can contribute significantly to the BPVE. However, we want to point out that in order to get a full understanding of the ballistic current and the BPVE, other scattering mechanisms such as defect scattering and electron-hole Coulomb scattering should also be taken into account.





As revealed by previous study, the shift current response can be strongly enhanced by modest changes to crystal structure or composition~\cite{gong2018phonon,wang2016substantial}.  Here, we extend this idea and explore the relation between the ballistic current and structure. We find that certain structures can greatly enhance the current response. To illustrate this point, we lift all constraints of BaTiO$_3$ and perform a full structural relaxation, so that the low-temperature rhombohedral phase is obtained. For this low-temperature structure, the corresponding ballistic current photovoltaic tensor is shown in Fig.~\ref{fig:response}(b). Its lineshape is dramatically different from that of the tetragonal phase (Fig.~\ref{fig:response}a), and the overall magnitude is much larger. Through a visual inspection of the two structures (the insets of Fig.~\ref{fig:response}(a) and (b)), we find a larger off-center displacement along the (111) direction in the rhombohedral structure and a smaller distortion along the (100) direction in the tetragonal phase. This could indicate a relation between the magnitudes of the current response and the structure distortion. Specifically, it could be that a larger extent of symmetry breaking will enhance the asymmetry of the momentum distribution, and the off-center displacement suggests that different parts of the Brillouin zone will not contribute to the ballistic current uniformly. A more quantitative investigation into the relationship between structure and the ballistic current will be the topic of our future study. For practical applications, however, this contrast between the ballistic current responses of rhombohedral and tetragonal \ch{BaTiO3} is very illuminating since it shows that a large part of the solar spectrum can be harvested by engineering the distortion via doping or external strain. 

In conclusion, based on the Kubo formula, we derived an expression for the phonon-assisted ballistic current, and we implement it into first-principles calculation. Taking \ch{BaTiO3} as an example, we demonstrated via first-principles calculations that the electron-phonon coupling is an important mechanism of the ballistic current and can contribute significantly to the BPVE. We showed that, similar to the shift current, the ballistic current is also very sensitive to structures; this reflects a promising possibility of material engineering to further harvest BPVE.

We acknowledge valuable discussions with Dr.\ Ruixiang Fei. This work has been supported by the Department of Energy Office of Basic Energy Sciences, under grant number DE-FG02-07ER46431. L.Z.T was supported by the Molecular Foundry, a DOE Office of Science User Facility funded by the Office of Science of the U.S. Department of Energy under Contract No. DE-AC02-05CH11231. The analytical derivation was carried out by L.Z.T and Z.D., and the numerical simulation was done by Z.D., A.M.S., and L.G.. Computational support is provided by the National Energy Research Scientific Computing (NERSC) Center of the
U.S. DOE.

\bibliography{BC}



\section*{Supplementary material}

# Ballistic Current From First Principle Calculations: Supplementary Information

Zhenbang Dai,[1, *] Aaron M. Schankler,[1, *] Lingyuan Gao,[1] Liang Z. Tan,[2] and Andrew M. Rappe[1]

[1]*Department of Chemistry, University of Pennsylvania,*
*Philadelphia, Pennsylvania 19104–6323, USA*
[2]*Molecular Foundry, Lawrence Berkeley National Laboratory, Berkeley, California 94720, USA*
(Dated: December 13, 2020)

**DETAILED DERIVATION OF EQ. 7**

We notice here that we are concerned with a semiconductor with a relatively large band-gap, so the electron population dictated by the Fermi distribution will be barely influenced by temperature at 300 K. Therefore, in order to simplify the derivation for this specific system, we will choose the real-time Green's function formalism in the following derivation, and we will show in the next section that the real-time Green's function formalism is equivalent to the more general imaginary-time Green's function formalism for systems with a wide band-gap.

By applying Feynman rule[1, 2] in the momentum-frequency space on FIG.1(a) in the main text, we can get:

$$\Delta\chi^{(2)\alpha\beta}(\omega) = \frac{1}{\hbar} \sum_{kk'cc'vv'\mu} \langle vk|P^\alpha|ck\rangle \langle c'k'|P^\beta|v'k'\rangle g^{cc'}_{\mu kk'} g^{v'v}_{\mu k'k}$$
$$\times -\frac{1}{\hbar^2} \int_{-\infty}^{+\infty} \int_{-\infty}^{+\infty} \frac{d\omega' d\omega''}{2\pi\ 2\pi} G_c(\omega',k) G_v(\omega'-\omega,k) G_{c'}(\omega'',k') G_{v'}(\omega''-\omega,k') D_\mu(\omega'-\omega'',k-k') \quad (S1)$$

where[2]

$$G_c(\omega,k) = \frac{1}{\omega - E_{ck}/\hbar + i0^+},$$
$$G_v(\omega,k) = \frac{1}{\omega - E_{vk}/\hbar - i0^+},$$
$$D_\mu(\omega,k) = (N_q+1)\left(\frac{1}{\omega-\omega_q+i0^+} - \frac{1}{\omega+\omega_q-i0^+}\right) + N_q\left(\frac{1}{\omega+\omega_q+i0^+} - \frac{1}{\omega-\omega_q-i0^+}\right), \quad (S2)$$

and $N_q$ is the phonon population, which is evaluated at 300 K in numerical evaluation. Note, all the symbols of wave vectors $k$ and $q$ denote their vector counterparts $\boldsymbol{k}$ and $\boldsymbol{q}$ for simplicity throughout the Supplementary Information unless otherwise specified.

Combining Eq. S1 and Eq. S2, we proceed by applying the residue theorem sequentially on $\omega'$ and $\omega''$, which will yield:

$$\Delta\chi^{(2)\alpha\beta}(\omega) = -\sum_{kk'cc'vv'\mu} \langle vk|P^\alpha|ck\rangle \langle c'k'|P^\beta|v'k'\rangle g^{cc'}_{\mu kk'} g^{v'v}_{\mu k'k}$$
$$\times \left\{(N_q+1)\frac{1}{E_{ck}-E_{vk}-\hbar\omega-i0^+} \frac{1}{E_{c'k'}-E_{v'k'}-\hbar\omega-i0^+}\right.$$
$$\times \left(\frac{1}{E_{c'k'}-E_{vk}+\hbar\omega_q-\hbar\omega-i0^+} + \frac{1}{E_{ck}-E_{v'k'}+\hbar\omega_q-\hbar\omega-i0^+}\right) + N_q[\omega_q \Leftrightarrow -\omega_q]\right\}$$
$$= -\sum_{kk'cc'vv'\mu} \langle vk|P^\alpha|ck\rangle \langle c'k'|P^\beta|v'k'\rangle g^{cc'}_{\mu kk'} g^{v'v}_{\mu k'k} I(c,c',v,v',k,k',\mu) \quad (S3)$$

Thus, the second-order carrier generation rate for an electron-hole pair can be written as:

$$\Delta\Gamma^{\alpha\beta}_{cv,k}(\omega) = -\frac{2}{\hbar} Im\left[\Delta\chi^{(2)\alpha\beta}_{cv,k}(\omega)\right]\left|\frac{e}{m\omega}E(\omega)\right|^2$$
$$= \frac{2}{\hbar}\left(\frac{eE(\omega)}{m\omega}\right)^2 Im\left[\sum_{k'c'v'\mu} \langle vk|P^\alpha|ck\rangle \langle c'k'|P^\beta|v'k'\rangle g^{cc'}_{\mu kk'} g^{v'v}_{\mu k'k} I(c,c',v,v',k,k',\mu)\right] \quad (S4)$$

Considering Eq. 6 and the fact that $I(c,c',v,v',-k,-k',\mu) = I(c,c',v,v',k,k',\mu)$ as the time-reversal symmetry makes $E(n,-k) = E(n,k)$ and $\omega_{-q} = \omega_q$, we are able to get the asymmetric carrier generation:

$$\Gamma^{\alpha\beta,asym}_{cv,k}(\omega) = \frac{1}{2}(\Delta\Gamma^{\alpha\beta}_{cv,k}(\omega) - \Delta\Gamma^{\alpha\beta}_{cv,-k}(\omega))$$
$$= \frac{2}{\hbar}\left(\frac{eE(\omega)}{m\omega}\right)^2 Im\left[\sum_{k'c'v'\mu} \langle vk|P^\alpha|ck\rangle \langle c'k'|P^\beta|v'k'\rangle g^{cc'}_{\mu kk'} g^{v'v}_{\mu k'k}\right] Re[I(c,c',v,v',k,k',\mu)] \quad (S5)$$

We can still move forward a little bit by substituting

$$\frac{1}{\omega \pm i0^+} = \mathcal{P}\frac{1}{\omega} \mp i\pi\delta(\omega) \quad (S6)$$

into $I(c,c',v,v',k,k',\mu)$. After collecting the terms that are real and satisfy the selection rule for optical transition, we can finally get Eq. 7 in the main text. If, however, the system possesses inversion symmetry, then the additional equalities will hold up to a phase:

$$\langle v,-k|P^\alpha|c,-k\rangle = -[\langle v,k|P^\alpha|c,k\rangle] \\ g^{nn'}_{\mu-k-k'} = g^{nn'}_{\mu k k'} \tag{S7}$$

and these equalities will make the asymmetric generation rate vanish.

## EQUIVALENCE BETWEEN THE REAL-TIME GREEN'S FUNCTION AND THE IMAGINARY-TIME GREEN'S FUNCTION

Here, we show that our way of using real-time Green's functions (GF) is equivalent to using imaginary-time Green's functions (Matsubara's functions, MF) for a wide-gap semiconductor. Essentially, we need to show the energy denominators $I(c,c',v,v',k,k',\mu)$ in Eq. S3 can also be obtained by MF formalism if there exists a wide band-gap. For this purpose, we ignore prefactors to simplify our proof.

In MF formalism, there will be a frequency sum after applying the Feynman rule:

$$I = -\sum_{nm} \frac{1}{i\omega_n - E_{ck}} \frac{1}{i\omega_n - i\omega - E_{vk}} \frac{1}{i\omega_m - E_{c'k'}} \frac{1}{i\omega_m - i\omega - E_{v'k'}} \frac{2\omega_q}{(i\omega_n - i\omega_m)^2 - \omega_q^2} \tag{S8}$$

where

$$\omega_n = (2n+1)\pi/\beta,\ \omega_m = (2m+1)\pi/\beta,\ \omega = 2l\pi/\beta, \\ n, m, l \in \mathbb{Z} \quad (\hbar = 1) \tag{S9}$$

First consider the summation over $\omega_m$:

$$I_m = \sum_m \frac{1}{i\omega_m - E_{c'k'}} \frac{1}{i\omega_m - i\omega - E_{v'k'}} \frac{2\omega_q}{(i\omega_m - i\omega_n)^2 - \omega_q^2}. \tag{S10}$$

Applying the frequency summation rule by [2], we obtain:

$$\begin{aligned} I_m = & - n_F(E_{c'k'}) \frac{1}{E_{c'k'} - i\omega - E_{v'k'}} \frac{2\omega_q}{(E_{c'k'} - i\omega_n)^2 - \omega_q^2} \\ & - n_F(i\omega + E_{v'k'}) \frac{1}{E_{v'k'} + i\omega - E_{c'k'}} \frac{2\omega_q}{(i\omega_n - i\omega - E_{v'k'})^2 - \omega_q^2} \\ & + n_F(i\omega_n + \omega_q) \frac{1}{i\omega_n + \omega_q - E_{c'k'}} \frac{1}{i\omega_n - i\omega + \omega_q - E_{v'k'}} \\ & - n_F(i\omega_n - \omega_q) \frac{1}{i\omega_n - \omega_q - E_{c'k'}} \frac{1}{i\omega_n - i\omega - \omega_q - E_{v'k'}} \end{aligned} \tag{S11}$$

Here, $n_F$ is Fermi-Dirac distribution, and we have shifted the Fermi level to 0 eV such that

$$n_F(z) = \frac{1}{e^{\beta z}+1},\ \beta = \frac{1}{k_B T} \tag{S12}$$

Since we are considering a wide-gap semiconductor at room temperature where $E_{ck} \gg 1/\beta > 0$ and $E_{vk} \ll -1/\beta < 0$,

$$\begin{aligned} n_F(E_{c'k'}) &\to 0 \\ n_F(i\omega + E_{v'k'}) &= \frac{1}{e^{\beta(i\omega+E_{v'k'})}+1} = \frac{1}{e^{i2l\pi}e^{\beta E_{v'k'}}+1} = \frac{1}{e^{\beta E_{v'k'}}+1} \to 1 \\ n_F(i\omega_n + \omega_q) &= \frac{1}{e^{\beta(i\omega_n+\omega_q)}+1} = \frac{1}{e^{i(2n+1)\pi}e^{\beta\omega_q}+1} = \frac{1}{1-e^{\beta\omega_q}} = -N_q \\ n_F(i\omega_n - \omega_q) &= \frac{1}{e^{\beta(i\omega_n-\omega_q)}+1} = \frac{1}{e^{i(2n+1)\pi}e^{-\beta\omega_q}+1} = \frac{1}{1-e^{-\beta\omega_q}} = 1 + \frac{1}{e^{\beta\omega_q}-1} = N_q + 1 \end{aligned} \tag{S13}$$

3

Therefore, $I_m$ can rewritten as:

$$I_m = -\frac{1}{E_{v'k'} + i\omega - E_{c'k'}} \frac{2\omega_q}{(i\omega_n - i\omega - E_{v'k'})^2 - \omega_q^2}$$
$$- N_q \frac{1}{i\omega_n + \omega_q - E_{c'k'}} \frac{1}{i\omega_n - i\omega + \omega_q - E_{v'k'}}$$
$$- (N_q + 1) \frac{1}{i\omega_n - \omega_q - E_{c'k'}} \frac{1}{i\omega_n - i\omega - \omega_q - E_{v'k'}} \quad (S14)$$

and the overall summation now is:

$$I = -\sum_n \frac{1}{i\omega_n - E_{ck}} \frac{1}{i\omega_n - i\omega - E_{vk}} I_m$$
$$= -\sum_n \frac{1}{i\omega_n - E_{ck}} \frac{1}{i\omega_n - i\omega - E_{vk}} \Bigg[ -\frac{1}{E_{v'k'} + i\omega - E_{c'k'}} \frac{2\omega_q}{(i\omega_n - i\omega - E_{v'k'})^2 - \omega_q^2}$$
$$- N_q \frac{1}{i\omega_n + \omega_q - E_{c'k'}} \frac{1}{i\omega_n - i\omega + \omega_q - E_{v'k'}} - (N_q + 1) \frac{1}{i\omega_n - \omega_q - E_{c'k'}} \frac{1}{i\omega_n - i\omega - \omega_q - E_{v'k'}} \Bigg] \quad (S15)$$

Follow the exact procedure and notice $|\omega_q| \ll |E_{ck}|$ (and $|\omega_q| \ll |E_{vk}|$), we can evaluate each term on the RHS of Eq. S15:

$$RHS1 = \frac{1}{i\omega + E_{v'k'} - E_{c'k'}} \frac{1}{i\omega + E_{vk} - E_{ck}} \frac{1}{i\omega + E_{v'k'} - E_{ck} + \omega_q}$$
$$- \frac{1}{i\omega + E_{v'k'} - E_{c'k'}} \frac{1}{i\omega + E_{vk} - E_{ck}} \frac{1}{i\omega + E_{v'k'} - E_{ck} - \omega_q}$$
$$RHS2 = -N_q \Bigg( \frac{1}{i\omega + E_{v'k'} - E_{c'k'}} \frac{1}{i\omega + E_{vk} - E_{ck}} \frac{1}{i\omega + E_{vk} - E_{c'k'} + \omega_q}$$
$$+ \frac{1}{i\omega + E_{v'k'} - E_{c'k'}} \frac{1}{i\omega + E_{vk} - E_{ck}} \frac{1}{i\omega + E_{v'k'} - E_{ck} - \omega_q} \Bigg)$$
$$RHS3 = -(N_q + 1) \Bigg( \frac{1}{i\omega + E_{v'k'} - E_{c'k'}} \frac{1}{i\omega + E_{vk} - E_{ck}} \frac{1}{i\omega + E_{vk} - E_{c'k'} - \omega_q}$$
$$+ \frac{1}{i\omega + E_{v'k'} - E_{c'k'}} \frac{1}{i\omega + E_{vk} - E_{ck}} \frac{1}{i\omega + E_{v'k'} - E_{ck} + \omega_q} \Bigg) \quad (S16)$$

then,

$$I = RHS1 + RHS2 + RHS3$$
$$= (N_q + 1) \frac{1}{E_{c'k'} - E_{v'k'} - i\omega} \frac{1}{E_{ck} - E_{vk} - i\omega}$$
$$\times \Bigg( \frac{1}{E_{ck} - E_{v'k'} - \omega_q - i\omega} + \frac{1}{E_{c'k'} - E_{vk} - \omega_q - i\omega} \Bigg) + N_q[\omega_q \Leftrightarrow -\omega_q] \quad (S17)$$

Therefore, after the analytical continuation, we can recover $I(c, c', v, v', k, k', \mu)$ in Eq. S1. The crucial condition for this equivalence is the material's wide band-gap, which will make $n_F(E_{ck}) \to 0$ and $n_F(E_{vk}) \to 1$.

## PROOF THAT OTHER DIAGRAMS DO NOT CONTRIBUTE TO THE ASYMMETRIC SCATTERING

Since the algebraic structure of FIG.1(b) and FIG.1(c) will be the same and they can be obtained from each by a relabeling c⇔v, it suffices to only prove that FIG.1(b) has no contribution to asymmetric carrier generation. Following the same procedure for FIG.1(a), we can get the second-order correction to carrier generation from FIG.1(b) as:

$$\Gamma^{\alpha\beta(b)}_{cv,k}(\omega) = \mathcal{N} Im \Bigg[ \sum_{k'c'v'\mu} \langle vk|P^\alpha|ck\rangle \langle ck|P^\beta|vk\rangle \left|g^{cc'}_{\mu kk'}\right|^2 I^{(b)}(c, c', v, k, k', \mu) \Bigg] \quad (S18)$$

Here, the additional superscript (b) indicates that it represents the contribution from FIG.1(b), and $\mathcal{N}$ is the prefactor, which is a real constant. $I^{(b)}(c, c', v, k, k', \mu)$ is the result after the frequency integrals of the Green's function. Similar to the $I(c, c', v, v', k, k', \mu)$ in Eq. S3, it is a function of the electronic energies and the phonon

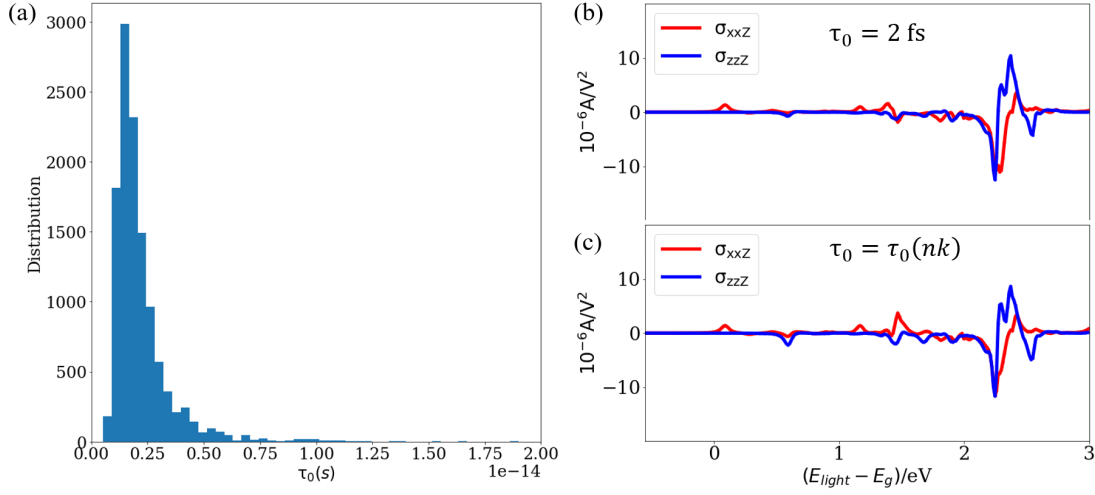

FIG. S1. (a) Distribution of the momentum relaxation time of different $(c, \mathbf{k})$. (b) The current response for constant relaxation-time approximation. (c) The current response for k-dependent relaxation-time.

energies, which will be unchanged under the change from k to -k. To proceed, we notice that $\Gamma_{cv,k}^{\alpha\beta(b)}(\omega)$ and $\Gamma_{cv,k}^{\beta\alpha(b)}(\omega)$ will be inseparable in experiments, and therefore they should always be considered simultaneously:

$$
\left[\Gamma_{cv,k}^{\alpha\beta(b)}(\omega) + \Gamma_{cv,k}^{\beta\alpha(b)}(\omega)\right] - \left[\Gamma_{cv,-k}^{\alpha\beta(b)}(\omega) + \Gamma_{cv,-k}^{\beta\alpha(b)}(\omega)\right]
$$
$$
= \mathcal{N} Im\Bigg\{ \sum_{k'c'v'\mu} \left|g_{\mu kk'}^{cc'}\right|^2 I^{(b)}(c,c',v,k,k',\mu)
$$
$$
\times \left[\langle vk|P^\alpha|ck\rangle\langle ck|P^\beta|vk\rangle + \langle vk|P^\beta|ck\rangle\langle ck|P^\alpha|vk\rangle - \langle vk|P^\alpha|ck\rangle^*\langle ck|P^\beta|vk\rangle^* - \langle vk|P^\beta|ck\rangle^*\langle ck|P^\alpha|vk\rangle^*\right]\Bigg\}
$$
$$
= 0 \tag{S19}
$$

Thus, we have proved that FIG.1(b) has no contribution to the asymmetric carrier generation, and so does FIG.1(c).

### ESTIMATION OF THE MOMENTUM RELAXATION TIME $\tau_0$

We use the as-generated electron-phonon coupling matrices for tetragonal phase BaTiO$_3$ to estimate the momentum relaxation time according to the formula in [3]. As can be seen in FiG. S1a, the statics of all the momentum relaxation times at different $(c, \mathbf{k})$ are very narrowly distributed and centered around 2 fs. So, we calculate the current response by taking $\tau_0 = 2$ fs (FIG. 2a and FIG. S1b), and we also compare it with the current response that takes the band/k-dependence of the lifetime into account (FIG. S1c). We found that the k-dependence of the lifetime barely influences the response lineshape and magnitude. Therefore, we approximate $\tau_0 = 2$ fs throughout this work, and similar approximation is well-adopted in many previous works, such as [4, 5].

### INFLUENCE OF TEMPERATURE

We also investigate the influence of temperature on the BPVE by calculating the total current for tetragonal BaTiO$_3$ at various temperatures. However, we would like to point out that the effect of temperature is complicated. For example, there will be thermal expansion, phase transition, change of magnitude of the polarization within the same phase, piezoelectric response to the polarization change, and so on, all of which will influence the phonon scattering and even the shift current. FIG. 2(a) and 2(b) in the main text show how the phase transition will influence the ballistic current, which could be perceived as one possible effect of temperature. Therefore, a comprehensive study of the effect of temperature goes beyond the scope of the present work, and our results should be viewed as a preliminary attempt to address one aspect of the temperature, which is its influence on the phonon population.

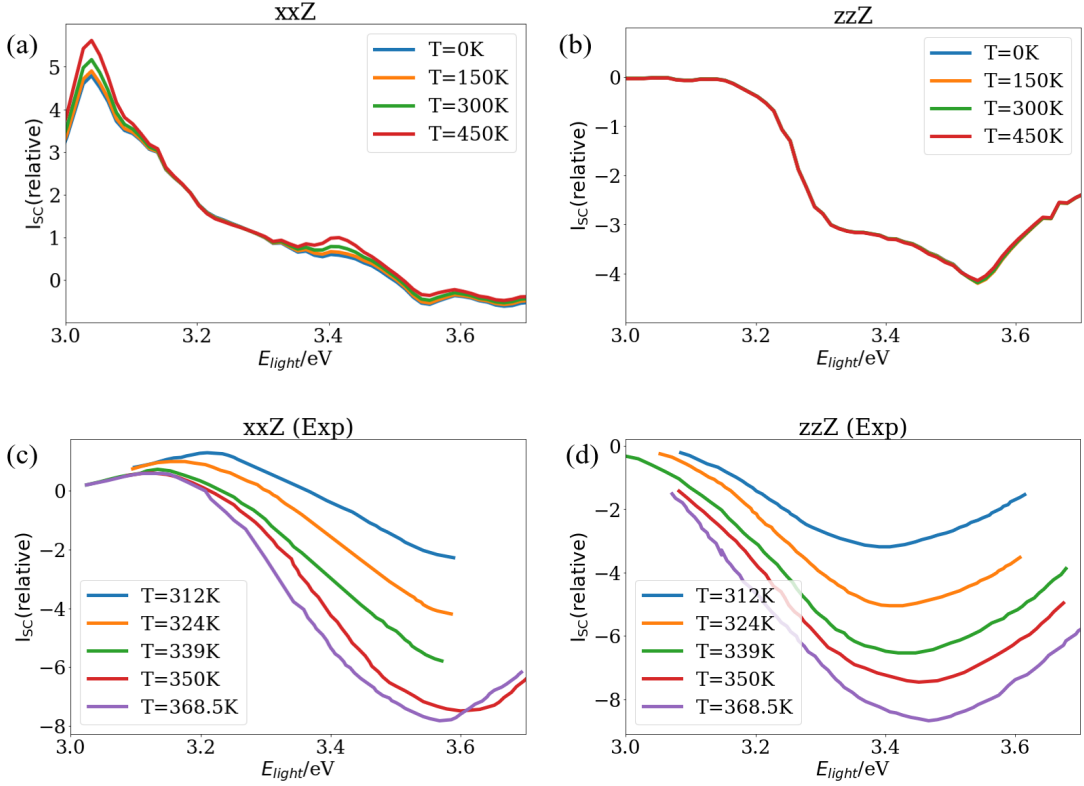

FIG. S2. Temperature influence on the ballistic current response of $BaTiO_3$. (a) and (b): Theoretical results of the total current at different temperatures by only considering the change of phonon population in the carrier generation rate. (c) and (d): Experimental results reproduced from [6]. The unit of the current is relative to $8 \times 10^{-12} A$.[6]

From FIG. S2(a) and S2(b), we can see that the temperature will have a larger influence on the xxZ component than the zzZ component, which indicates that some higher-energy phonon modes contribute to the zzZ component in the shown frequency range because they are less influenced by the temperature. And a general trend from FIG. S2(a) is that the lower the temperature is, the smaller magnitude the total current will be due to the smaller ballistic current with less phonon population, in contrast to shift current which will not be influenced by temperature for a semiconductor. However, the experimental spectra show a larger temperature dependence as shown in FIG. S2(c) and S2(d)[6], which again illustrates that the effect of temperature can be complicated, and more factors mentioned above need to be taken into account in order to fully understand it.

## GENERALIZED NEWTON-COTES METHOD

Numerically, for principal part (PP) integrals like Eq. 8, the common practice is to add a small pure imaginary number in the denominator to 'smear' it, but it usually requires a very dense $k$-grid to get a converged result. However, since our final expression Eq. 1 has a double sum over $k$, it is computationally very expensive both to get the required ingredients and to perform the summation itself for a denser $k$-grid, which will make such calculation prohibitive. To circumvent this problem, we devised a new numerical PP integral scheme inspired by [7]. For the most general case, we are dealing with a three-dimensional integral in the form of:

$$I = \mathcal{P} \int_V d^3k \frac{f(\vec{k})}{g(\vec{k}) - p} \quad (S20)$$

where p is a parameter and V represents the spatial region where the integration is performed. Our scheme is to split the overall region into several regions that are small enough:

$$I = \sum_{\Delta V} I(\Delta V) = \sum_{\Delta V} \mathcal{P} \int_{\Delta V} d^3k \frac{f(\vec{k})}{g(\vec{k}) - p}. \quad (S21)$$

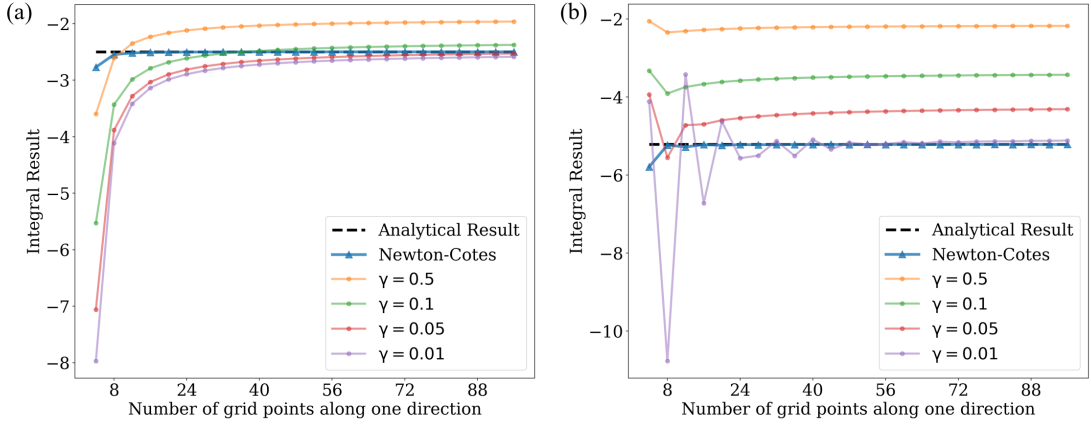

FIG. S3. The comparison between the generalize Newton-Cotes method and the smearing method for integrals in Eq. S25. $\gamma$ is the value of the small imaginary number in the smearing method. (a) $p = 0.6$. (b) $p = 0.4$. The smearing method requires a very dense grid to give well-converged results, but the GNC method can yield converged results even for a very small number of grid points. In addition, the GNC method is not sensitive to the position of the poles as it can give fast convergence for both $p = 0.6$ and $p = 0.4$. For the smearing method, however, even though when $p = 0.6$, various smearing values can yield convergent result given denser grids as shown in (a), only $\gamma = 0.01$ will give rise to convergent result when $p = 0.4$, which can be seen in (b).

Since each integration region is small, we are allowed to perform the multi-variate Taylor expansion on $f(\vec{k})$ and $g(\vec{k})$ without loss too much of the accuracy:

$$I(\Delta V) \approx \mathcal{P} \int_{\Delta V} d^3 k \frac{f(\vec{k_0}) + \frac{\partial f}{\partial k_x}\big|_{\vec{k_0}}(k_x - k_{x0}) + \frac{\partial f}{\partial k_y}\big|_{\vec{k_0}}(k_y - k_{y0}) + \frac{\partial f}{\partial k_z}\big|_{\vec{k_0}}(k_z - k_{z0})}{g(\vec{k_0}) + \frac{\partial g}{\partial k_x}\big|_{\vec{k_0}}(k_x - k_{x0}) + \frac{\partial g}{\partial k_y}\big|_{\vec{k_0}}(k_y - k_{y0}) + \frac{\partial g}{\partial k_z}\big|_{\vec{k_0}}(k_z - k_{z0}) - p} \tag{S22}$$

In this small region $\Delta V$, $k_{x0}$, $k_{y0}$, $k_{z0}$, $f(\vec{k_0})$, $g(\vec{k_0})$, and the partial derivatives calculated from two-points formula can be taken as constants, so we are left with two types simple integrals, which has analytical forms for a cuboid region:

$$\mathcal{P} \int_{x_1}^{x_2} \int_{y_1}^{y_2} \int_{z_1}^{z_2} dxdydz \frac{1}{Ax + By + Cz - p'} \tag{S23}$$

$$\mathcal{P} \int_{x_1}^{x_2} \int_{y_1}^{y_2} \int_{z_1}^{z_2} dxdydz \frac{x}{Ax + By + Cz - p'} \tag{S24}$$

where A, B, C and p' are constants. By doing these integrations analytically, we can get the integral for the volume element $\Delta V$, of which the summation would yield the the overall the value of the integral. We can see that this method is fact a generalized Newton-Cotes(GNC) method and the analytical integration over the poles will be expected to eliminate the numerical instability.

We test the GNC method against the smearing method by doing both on a simple integral with two variables which can be done analytically:

$$f(p) = \mathcal{P} \int_{-0.5}^{0.5} \int_{-0.5}^{0.5} dxdy \frac{1}{x^2 + y^2 - p} \tag{S25}$$

and the results are shown in FIG.S3. As expected, the traditional smearing method requires a very dense grid to give well-converged results, but the GNC method can yield converged results even for a fairly small number of grid points. Another advantage of the GNC method over the smearing method is its insensitivity to the position of the poles. In FIG.S3(a) where $p = 0.6$, various smearing values can yield convergent result given denser grids, whereas when $p = 0.4$ as shown in FIG.S3(b), only $\gamma = 0.01$ will give rise to convergent result. Notably, however, the GNC method can give fast convergence for small number of grid points in both cases. Therefore, in our calculations, we choose the GNC method to evaluate Eq. 7 and Eq. 8.

---

* These authors contributed equally


\end{document}